\documentclass[onecolumn,preprint,superscriptaddress]{revtex4-1}
\usepackage{bibentry,natbib}
\usepackage{latexsym,bm}
\usepackage{graphicx}
\usepackage{rotating}
\usepackage{amsmath}
\usepackage{rotating}
\usepackage{color}
\usepackage[colorlinks,citecolor=red]{hyperref}
\usepackage{subfigure}
\usepackage{ulem}
\DeclareGraphicsExtensions{.jpg}

\begin{document}
\title{Effects of $p$-wave Interactions on Borromean Efimov Trimers in Heavy-Light Fermi Systems}
\author{Cai-Yun Zhao}
\affiliation{State Key Laboratory of Magnetic Resonance and Atomic and Molecular Physics, Wuhan Institute of Physics and Mathematics, Innovation Academy for Precision Measurement Science and Technology, Chinese Academy of Sciences, Wuhan 430071, P. R. China}
\affiliation{ University of Chinese Academy of Sciences, 100049, Beijing, P. R. China}
\author{Hui-Li Han}
\email{huilihan@wipm.ac.cn}
\affiliation{State Key Laboratory of Magnetic Resonance and Atomic and Molecular Physics, Wuhan Institute of Physics and Mathematics, Innovation Academy for Precision Measurement Science and Technology, Chinese Academy of Sciences, Wuhan 430071, P. R. China}
\author{Ting-Yun Shi}
\affiliation{State Key Laboratory of Magnetic Resonance and Atomic and Molecular Physics, Wuhan Institute of Physics and Mathematics, Innovation Academy for Precision Measurement Science and Technology, Chinese Academy of Sciences, Wuhan 430071, P. R. China}
\date{\today}

\begin{abstract}
We investigate the effects of $p$-wave interactions on Efimov trimers in systems comprising two identical heavy fermions and a light particle, with mass ratios larger than $13.6$. Our focus lies on the borromean regime where the ground-state trimer exists in the absence of dimers. Using pair-wise Lennard-Jones potentials and concentrating on the $L^{\pi} = 1^{-}$ symmetry, we explore the critical value of the interspecies $s$-wave scattering length $a_{c}$ at which the borromean state appears in several two-component particle systems. Our exploration encompasses the universal properties of $a_{c}$ and the influence of $p$-wave fermion-fermion interactions on its value. We find that, in the absence of $p$-wave fermion-fermion interactions, $a_{c}$ is determined universally by the van der Waals radius and mass ratio. However, the introduction of $p$-wave fermion-fermion interactions unveiled a departure from this universality. Our calculations show that the critical interspecies scattering length $a_{c}$ now depends on the details of the fermion-fermion $p$-wave interaction. And, the presence of $p$-wave fermion-fermion interactions favors the formation of the borromean state. Additionally, our investigation reveals that Efimov effect in the $1^{-}$ symmetry persist even when the fermion-fermion interaction reaches the $p$-wave unitary limit.
\end{abstract}
\maketitle
\section{Introduction}
Universality and scaling symmetry in few-body systems have garnered considerable attention\;\cite{hwhammer2006,Greene2017Aug,Naidon2017}. Among the various phenomena, the Efimov effects, initially predicted by Efimov in 1970\;\cite{Efimov1970}, have attracted broad interest in studies in atomic and nuclear physics\;\cite{hwhammer2006,Greene2017Aug,Naidon2017}. The most striking manifestation of the Efimov effect is the possibility of infinite three-body bound states when the two-body $s$-wave scattering length $a$ tends to infinity. These three-body bound states, called Efimov states, are not affected by the details of the interaction and follow the discrete symmetry scaling law \;$E_n=\lambda^2E_{n+1}$, where $\lambda=e^{\pi/ s_0}$ is the scaling constant and $s_0$ is a universal parameter\;\cite{hwhammer2006,Greene2017Aug,Naidon2017,Efimov1970}.

The Efimov effect has been extensively studied in bosonic systems both theoretically and experimentally over the past decades\;\cite{DIncao2006mass,Zaccanti2009Cs,Huang2014Cs,Gross2010Li7,Kunitski2015He,Pires2014cslia,UlmanisKuhnle2016csli,Johansen2017csli}. In principle, the three-body parameter could be defined in terms of any observable related to Efimov physics. One of the definitions is the value of $a_{-}^{(1)}$ at which the first Efimov resonance appears in the three-body recombination\;\cite{hwhammer2006,Naidon2017}. Experiments with
alkali atoms have observed a universal value for the first Efimov resonance $a_{\scriptscriptstyle-}^{\scriptscriptstyle(1)}$, which is universally determined by the van der Waals (vdW) length $r_{\text{vdW}}$: $a_{\scriptscriptstyle-}^{\scriptscriptstyle (1)} = -( 9.1\pm 1.5)\, r_{\text{vdW,HL}}\;[r_{\text{vdW,HL}} = (2 \mu_{\scriptscriptstyle 2b} C_{6})^{1/4}/2]$\;\cite{Naidon2017,jiawangPRL2012} in the homonuclear systems. This universality arises from a universal effective barrier in the three-body potentials that prevents the three particles from simultaneously approaching each other\;\cite{Sorensen2012,jiawangPRL2012,Huang2014Li6,Naidon2014PRL}. For systems composed of two identical fermions (F) and one distinguishable particle (A), where each fermion interacts with another particle near an $s$-wave resonance denoted as $a_{s}$, which is the most important quantity to describe low-energy collision processes, while interactions between the two identical fermions occur in the often-disregarded $p$-wave channel. Notably, the emergence of the Efimov effect within F-F-A systems is predicted in the $p$-wave channel when the mass ratio exceeds a critical value of 13.601\;\cite{hwhammer2006,Helfrich2011}. Intriguingly, below this critical mass ratio, these systems can still support universal Kartavtsev-Malykh trimers (KM trimers)\;\cite{KartavtsevAtomicNuclei} and crossover trimers\;\cite{Endo2012Dec} in the limit of the infinite and finite positive $s$-wave scattering length, respectively. The characteristics of these trimers have been exhaustively studied within the framework of the zero-range approximation while neglecting $p$-wave fermion-fermion interactions\;\cite{hwhammer2006,Helfrich2011,KartavtsevAtomicNuclei}. Yet, the critical value of the interspecies scattering length, denoted as $a_c$, signifying the appearance of the first $p$-wave Efimov state, along with its potential vdW universality similar to the $s$-wave Efimov effect, remains largely unknown. Moreover, in a physical system comprising two identical fermions and a distinguishable particle, the coexistence of both s- and p-wave interactions may takes on notable significance\;\cite{Qin2016Dec}. It is therefore of interest to explore the implications of the p-wave fermion-fermion interaction on the universality of F-F-A systems.

It is known that Efimov effects do not occur in systems with resonant $p$-wave interactions\;\cite{Nishida2012Jul}. Recent studies by Chen and Greene\;\cite{Chen2023Mar} also revealed that no true Efimov effect appears in systems comprising three equal-mass fermions with mixed $s$-wave and $p$-wave interactions. However, in these cases, one or two $p$-wave unitary channels emerge at unitary. Moreover, a recent study\;\cite{1021468SciPostPhys} indicates the existence of $p$-wave induced trimers in mass-imbalanced F-F-A systems when a sufficiently large and positive scattering length is considered, along with the presence of $p$-wave interactions between two fermions. In systems with a mass ratio greater than 13.6, which support Efimov effect in $1^{-}$ channel, additional $p$-wave unitary channels are anticipated when fermion-fermion interactions are taken into account. This raises the question of whether the presence of $p$-wave fermion-fermion interactions alters the nature of Efimov effect in $1^{-}$ channel in F-F-A systems and what impact these interactions have on the Efimov three-body parameter.

To address these questions, we consider a system consisting of two heavy identical fermions of mass $M$ and one light particle with mass $m$, interacting via Lennard-Jones potentials that effectively mimic the finite range and repulsive core of realistic interactions. We investigate the critical value of the interspecies $s$-wave scattering length $a_{c}$ at which the borromean state with $L^{\pi}=1^{-}$ symmetry appears  for seven experimental systems of interest with mass ratios $M/m$ ranging from 6 to 30. The universal properties of $a_{c}$ and the influence of $p$-wave fermion-fermion interactions on its value are studied within the adiabatic hyperspherical representation.

This paper is organized as follows:
In Sec. II, we present our calculation method and all necessary formulas for calculations.
In Sec. III, we discuss the results and emphasize the significant role of $p$-wave interactions in F-F-A systems.
Finally, we conclude and summarize our work in Sec. IV.
Throughout the paper, atomic units are employed unless stated otherwise.
\section{Theoretical formalism}
We consider the general problem of two identical fermions of mass M and one particle of mass m, all three interacting with each other through finite-range interaction. The two fermions are denoted as particles 1 and 2, and the third particle as 3. The distance between particles $i$ and $j$ are labeled as $r_{ij}$.
In the frame of hyperspherical coordinates, the three-body Schr\"{o}dinger equation describing the relative motion of system can be expressed in terms of $R$ that describe the global size and hyperangles $ \Omega$ which include two hyperangles $ \theta $ and $\phi$ that describe the shape of the three-body system, and the three usual Euler angles $ \alpha,\beta,\gamma $ that describe the rigid-body rotations. After rescaled   by $\psi_{\upsilon'}(R;\Omega)=\Psi_{\upsilon'}(R;\Omega)R^{5/2}\sin\phi\cos\phi$,  the three-body equation for the relative motion can be written as:
\begin{equation}
\label{1}
\bigg[-\frac{1}{2\mu}\frac{d^2}{dR^2}+\bigg(\frac{\Lambda^2-\frac{1}{4}}{2\mu R^2}
+V(R;\theta,\phi)\bigg)\bigg]\psi_{\upsilon'}(R;\Omega) =E\psi_{\upsilon'}(R;\Omega)\,,
\end{equation}
where $\Lambda^2$ is the squared "grand angular momentum operator", whose expression is given in Ref\;\cite{CDLIN1995}.
$\mu=\sqrt{\frac{m_1 m_2 m_3 }{m_1+m_2+m_3}}$ is three-body reduce mass, where $1$, $2$ and $3$ refer to the fermion, fermion and the other atom respectively.
The total potential $V(R;\theta,\phi)$ in Eq.(\ref{1}) is defined as the pairwise sum of the two-body interactions with
\begin{equation}
\label{2}
V(R;\theta,\phi) = \upsilon(r_{12})+\upsilon(r_{23})+\upsilon(r_{31})\,
\end{equation}
and without
\begin{equation}
\label{222}
V(R;\theta,\phi) = \upsilon(r_{23})+\upsilon(r_{31})\,
\end{equation}
fermion-fermion $p$-wave interaction respectively.

The adiabatic Hamiltonian, containing all angular dependence and interactions, is defined as
\begin{equation}
H_{ad}(R,\Omega)=\left[\frac{\Lambda^2-1/4}{2\mu R^2}+V(R,\theta,\phi)\right].
\end{equation}
The adiabatic potentials $U_v(R)$ and the corresponding channel functions $\Phi_\nu(R,\Omega)$ at the fixed $R$ are obtained by solving the following equation:
\begin{align}
H_{ad}(R,\Omega)\Phi_\nu(R,\Omega)=U_\nu\Phi_\nu(R,\Omega),
\end{align}
whose solution depend parametrically on $R$.
The effective adiabatic potential $W_{v}$ in channel $\nu$ is:
\begin{align}
W_{\nu}(R)\equiv U_{\nu}(R)-\frac{1}{2\nu} Q_{\nu\nu}(R)\,.
\end{align}
The nonadiabatic coupling matrices $P_{\mu\nu}(R)$ and $Q_{\mu\nu}(R)$ are defined as
\begin{align}
\label{29}
P_{\mu\nu}(R)=\int d\Omega \Phi_{\mu}(R;\Omega)^{*}\frac{\partial}{\partial R}\Phi_{\nu}(R;\Omega)\,,
\end{align}
and
\begin{align}
\label{30}
Q_{\mu\nu}(R) = \int d\Omega \Phi_{\mu}(R;\Omega)^{*}\frac{\partial^{2}}{\partial R^{2}}\Phi_{\nu}(R;\Omega)\,.
\end{align}
The hyper-radial equation is solved by the slow variable discretization (SVD) method~\cite{Tolstikhin1996} combined with the finite-element method with discrete variable representation (FEM-DVR)\;\cite{Wumengshan2016,Rescigno2000}. We divide the hyperradial range into $n_{e}$ elements by grid points $R_{i}( i=0,1,...,n_{e} )$. Each finite-element is further subdivided by $M$ Gauss quadrature points. After rewriting and renumbering the basis set constructed by the Gauss-Lobatto shape function into a new basis set $\{\chi_{i}\}$, the total wave function $\psi_{l}(R,\phi,\theta)$ is expanded in terms of basis functions $\chi_{i}$ and the channel functions $\Phi_{\nu}(R_{i},\phi,\theta)$ as
\begin{equation}
\psi_{l}(R,\phi,\theta)=\sum_{i=1}^{N_{\scriptscriptstyle{\textsl{DVR}}}}\sum_{\nu=1}^{N_{\scriptscriptstyle{\textsl{chan}}}}C^{l}_{i\nu}\chi_{i}(R)\Phi_{\nu}(R_{i},\phi,\theta)\,,
\label{eq-hh}
\end{equation}
where $N_{\scriptscriptstyle{\textsl{DVR}}}=n_{e}\times(M-1)-1$ is total number of FEM-DVR basis and $N_{\scriptscriptstyle{\textsl{chan}}}$ is the number of the channel functions included. Using the orthogonality of FEM-DVR basis and channel functions, insertion of $\psi_{l}(R,\phi,\theta)$ into the three-body Schr\"{o}dinger equation Eq.~(\ref{1}) results in the standard algebraic eigenvalue problem for the coefficients $C^{l}_{i\nu}$
\begin{equation}
\sum_{i'=1}^{N_{\scriptscriptstyle{\textsl{DVR}}}}\sum_{\nu'=1}^{N_{\scriptscriptstyle{\textsl{chan}}}}\mathcal{T}_{ii'}\mathcal{O}_{i\nu,i'\nu'}C^{l}_{i'\nu'}+U_{\nu}({R}_{i})C^{l}_{i\nu}=E^{l}C^{l}_{i\nu}\,,
\end{equation}
where
\begin{equation}
\mathcal{T}_{ii'}=\frac{1}{2\mu}\int_{R_{\scriptscriptstyle{\textsl{0}}}}^{R_{\scriptscriptstyle{\textsl{max}}}}\frac{\mathrm{d}}{\mathrm{d}R}\chi_{i}(R)\frac{\mathrm{d}}{\mathrm{d}R}\chi_{i'}(R)\mathrm{d}R
\end{equation}
are the kinetic energy matrix elements with $R_{\scriptscriptstyle{\textsl{0}}}$ and $R_{\scriptscriptstyle{\textsl{max}}}$ are the boundaries of the calculated box, and
\begin{equation}
\mathcal{O}_{i\nu,i'\nu'}=\langle\Phi_{\nu}(R_{i},\phi,\theta)|\Phi_{\nu'}(R_{i'},\phi,\theta)\rangle
\end{equation}
are the overlap matrix elements between the adiabatic channels defined at different quadrature points. Based on the properties of the Lobatto shape functions, the matrix of the local potential-energy operator in FEM-DVR has a diagonal representation with matrix element values equal to potential values at grids points. More importantly, the kinetic energy matrix $\mathcal{T}$ is a band matrix, so we just need to calculate the overlap matrix $\mathcal{O}_{i\nu,i'\nu'}$ at adjacent intervals. This greatly reduces the computational efforts in calculating the overlap matrix elements.
The sizes of basis sets used in the $\phi$ and $\theta$ directions are $N_{\phi}=388$ and $N_{\theta}=212$ to make the hyperspherical potential curves have at least 5 significant digits. For the hyper-radial equation, we use the basis sets are: $N_{\scriptscriptstyle{\textsl{DVR}}}=500$ and $N_{\scriptscriptstyle{\textsl{chan}}}=10$ to ensure the convergence of three-body energy.
\section{Results and discussion}
\subsection{two-body system}
To model the interactions between two atoms, we use the Lennard-Jones potential,
 \begin{equation}
\label{39}
\upsilon(r_{ij})= -\frac{C_{6,ij}}{r^6_{ij}}\bigg[1-\frac{1}{2}(\frac{\gamma_{ij}}{r_{ij}})^6\bigg]\,,
\end{equation}
which has been widely used in the literature and is an excellent model potential for exploring vdW universality in Efimov physics\;\cite{yujunwangPRl2012,Ulmanis2016cslia,Wang201211,jiawangPRL2012,Naidon20148}. The Lennard-Jones potential has the vdW length and can avoid numerical difficulties in the short range.\;\cite{jiawangPRL2012}.
 The parameter $\gamma_{ij}$ in this model potential can be adjusted to give the desired scattering length and number of bound states.
The values of $C_6$ between two atoms adopted here are listed in Table \ref{t1}.
By adjusting the parameter $\gamma_{FF}$, we can study the influence of identical fermion interaction near or away from the $p$-wave resonance. Table \ref{t1} also lists the mass, two-body vdW dispersion coefficients, vdW length $r_{\text{vdW,FF/FA}}$, and Efimov scaling constant $s_{0}$ of the seven F-F-A systems.

 The $p$-wave interactions are described at low energy by two parameters: the $p$-wave scattering volume and the $p$-wave effective range. For low-energy $p$-wave scattering of a short-range potential, the variation of the phase shift $\delta_1(k)$ with energy $E=k^2/(2\mu)$ can be expressed in terms of only two parameters $V_p$ and $r_p$
 \begin{equation}
\label{399}
k^3 \delta_1(k)\rightarrow -\frac{1}{V_p}+\frac{1}{2}r_pk^2+\textsl{O}(k^4)\,,
\end{equation}
where $\delta_1$ is the $p$-wave phase shift, $V_p$ is the $p$-wave scattering volume, and $r_p$ is the $p$-wave effective range\;\cite{Madsen2002}. The units of the $p$-wave scattering volume and effective range
are (length)$^3$ and 1/(length), respectively.



\begin{table}
    \caption    { The mass, two-body vdW dispersion coefficients, Efimov scaling constant $s_{0}$, and $r_{\text{vdW,FF/FA}}$ of the seven FFA systems. }
    \label{t1}
    \renewcommand{\arraystretch}{0.8}
    \centering
    \renewcommand\arraystretch{0.8}
    \begin{tabular}{lccccccccc}
       \hline
       \hline
    $system( {\scriptstyle \textsl{F-F-A}})$&$M$&$m$ &$M/m$ &$C_{\scriptscriptstyle\textsl{6,FF}}$&$C_{\scriptscriptstyle\textsl{6,FA}}$&$s_{0}$ &$r_{\text{vdW,FF}}$ &$r_{\text{vdW,FA}}$  \\
       \hline
    $^{40}$K\,-$^{40}$K\,-$^{6}$Li      &$39.964$ &$6.015$ &$6.644$  &$3905^{a}$&$2322^c$ &$-$ &$65.022$ &$40.782$    \\
    $^{87}$Sr\,-$^{87}$Sr\,-$^{7}$Li    &$86.909$ &$7.016$ &$12.387$ &$3250^a$ &$2070^a$ &$-$ &$75.312$ &$41.832$    \\
    $^{87}$Sr\,-$^{87}$Sr\,-$^{6}$Li    &$86.909$ &$6.015$ &$14.448$ &$3250^a$ &$2070^a$ &$0.399$ &$75.312$ &$40.360$   \\
    $^{137}$Ba\,-$^{137}$Ba\,-$^{7}$Li  &$136.906$&$7.016$ &$19.513$ &$5160^b$ &$2637^c$ &$1.044$ &$94.717$ &$44.750$  \\
    $^{137}$Ba\,-$^{137}$Ba\,-$^{6}$Li  &$136.906$&$6.015$ &$22.760$ &$5160^b$ &$2637^c$ &$1.291$ &$94.717$ &$43.136$   \\
    $^{167}$Er\,-$^{167}$Er\,-$^{6}$Li  &$166.932$&$6.015$ &$27.752$ &$1723^d$ &$1508^e$ &$1.593$ &$75.660$ &$38.856$   \\
    $^{171}$Yb\,-$^{171}$Yb\,-$^{6}$Li  &$170.936$&$6.015$ &$28.418$ &$1909^f$ &$1606^f$ &$1.628$ &$78.085$ &$38.187$   \\
    \hline
    \hline
    \end{tabular}
    \begin{flushleft}
    $^{a}$From\cite{Mitroy2003NOV},
    $^{b}$From\cite{2006High},
    $^{c}$From\cite{2010Electric},
    $^{d}$From\cite{0Quantum},
    $^{e}$From\cite{PhysRevA92022708},
    $^{f}$From\cite{C6RbYbYbLi}.
    \end{flushleft}
    \end{table}

\subsection{Without Fermion-Fermion interaction}
\begin{figure}
 \centering
   \includegraphics[width=0.5\linewidth]{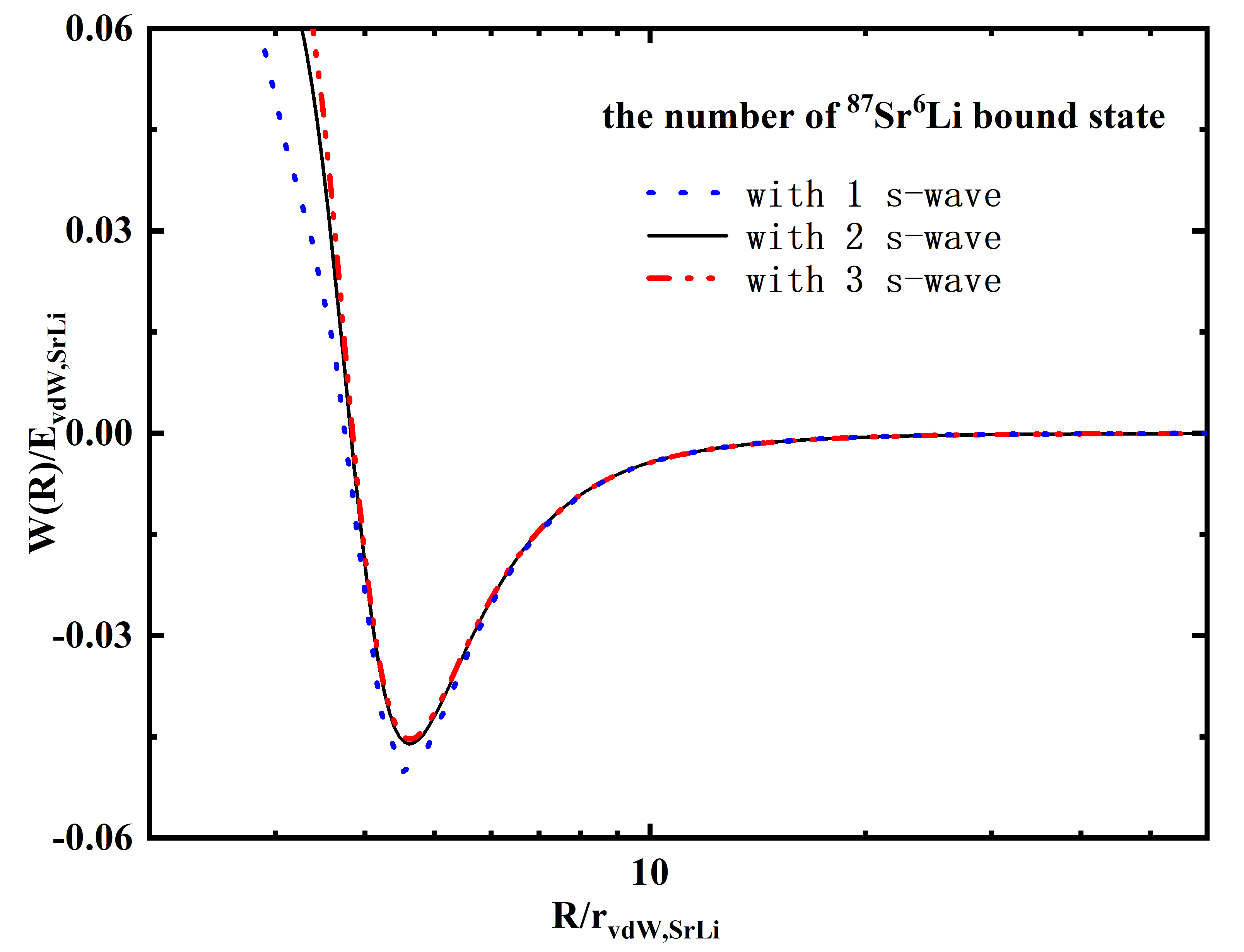}
\caption{The adiabatic potentials $W(R)$ with $L^{\pi} = 1^{-}$ symmetry for the $^{87}$Sr-$^{87}$Sr-$^{6}$Li system at $a_{\text{SrLi}}\rightarrow\infty$, obtained using the Lennard-Jones potential with different numbers of $s$-wave Sr-Li bound states.}
   \label{fig1}
\end{figure}

In this section, we focus on the regime where the interaction between the fermions can be neglected. In this situation, the Efimov effect only comes into play for a mass ratio $M/m \geq 13.6$. It is known that the universality of the three-body paparameter $a^{1}_{-}$ in s-wave Efimov effects arises from a universal effective barrier present in the three-body potentials. Our objective here is to explore the universality of $a_{c}$, concerning the emergence of the borromean state with symmetry $L^{\pi} = 1^{-}$. To achieve this, we present a series of the lowest adiabatic potentials, denoted as $W(R)$, for $^{87}$Sr-$^{87}$Sr-$^{6}$Li as $a_{SrLi}$ approaches infinity. These potentials have been computed at different $s$-wave poles within the Sr-Li potential, as depicted in Figure \ref{fig1}. It is shown that the effective adiabatic potentials converge to a single universal curve, indicating that the three-body parameter is universally independent of the short-range details of two-body potentials[ except for the van der waals length], similar to the efimov effect in the $0^{+}$ symmetry. All potentials
approach the universal form $-( s^2_{0} + 1/4 )/2 \mu R^2$ with $s_{0}=0.399$ when $R$ exceeds 10 times the van der Waals radius of the Sr-Li system, denoted as $r_{\text{vdw,SrLi}}$. Notably, they also show a repulsive wall at approximately $R= 3\; r_{\text{vdw,SrLi}}$, which prevents atoms from probing the small $R$ region representing the hallmark of van der Waals universality\cite{jiawangPRL2012,Chen2022Jan}. This repulsive feature remains consistent across varying numbers of two-body bound states, further emphasizing its independence from such details.

\begin{figure}
 \centering
   \includegraphics[width=0.5\linewidth]{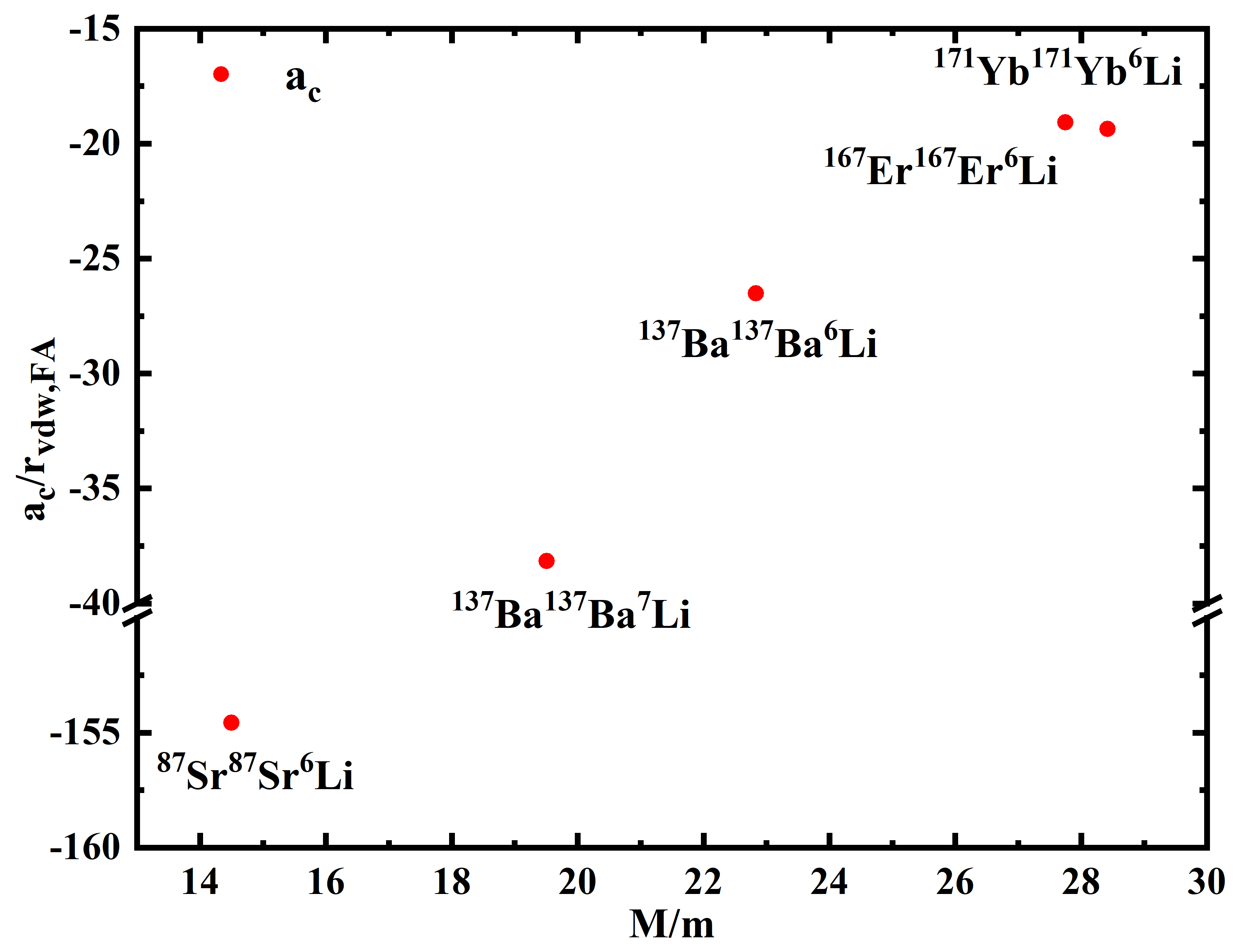}
\caption{The critical scattering length $a_{c}$, scaled by the van der Waals radius $r_{\text{vdw,FA}}$ of F-A , for the appearance of borromean states as a function of mass ratio for $L^{\pi} = 1^{-}$ symmetry without fermion-fermion interaction.}
   \label{fig2}
\end{figure}

\begin{figure}[htbp]
\centering
\includegraphics[width=0.5\linewidth]{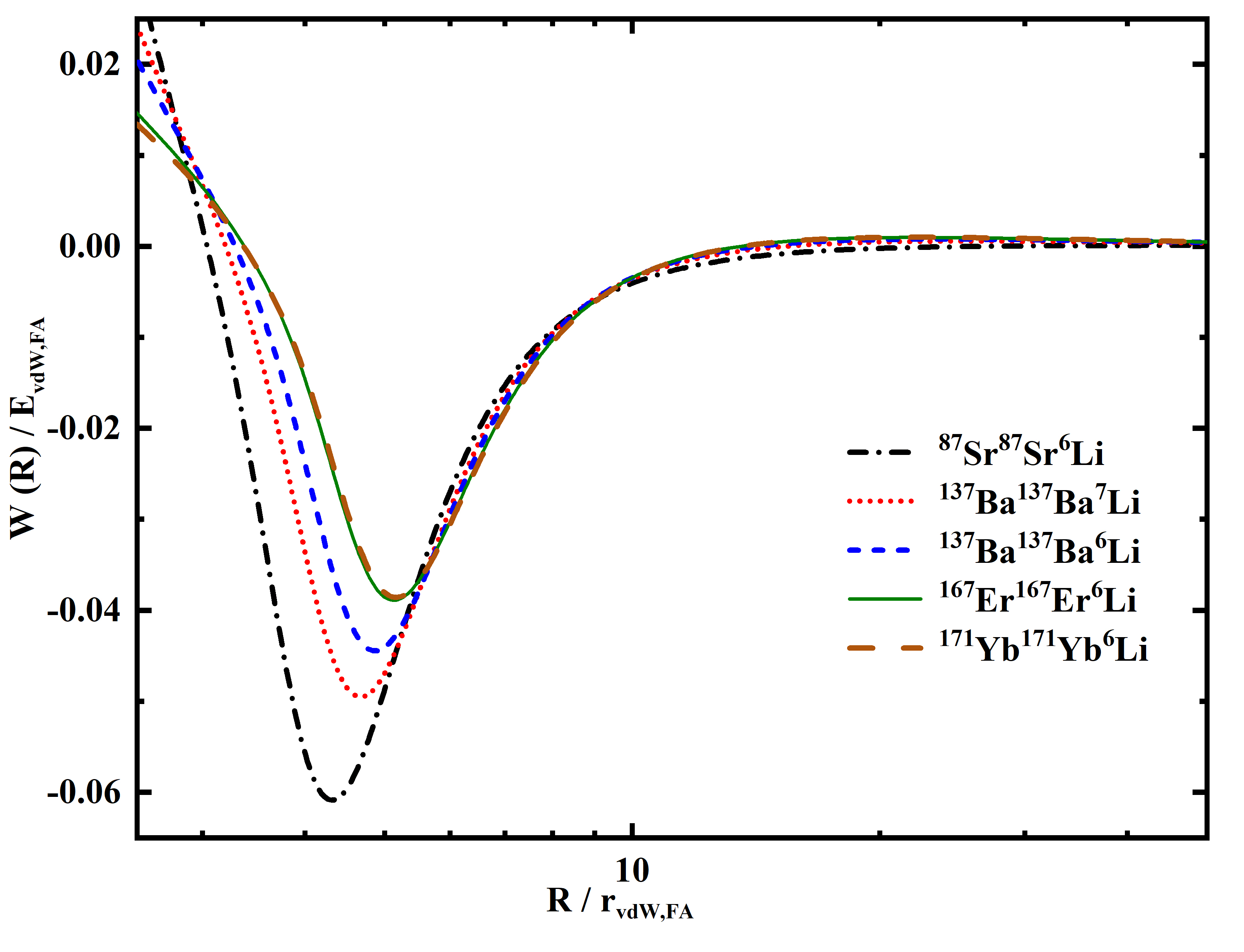}
\caption{(Color online)The adiabatic potentials $W(R)$ with $L^{\pi} = 1^{-}$ symmetry for F-F-A systems. The fermion-atom interaction is fixed at the critical value $a_{c}$, at which the $p$-wave borromean state appears.}
 \label{fig3}
\end{figure}

To further investigate the universality of the critical value $a_{c}$, we perform calculations across various systems to identify the specific values of $a_{c}$ required to observe the borromean state with $L^\pi=1^{-}$ symmetry without $p$-wave fermion-fermion interaction. As expected, we found no evidence of a three-body borromean state in the $^{87}$Sr,-$^{87}$Sr,-$^{7}$Li and $^{40}$K,-$^{40}$K,-$^{6}$Li systems, both of which possess mass ratios below the critical threshold of $13.6$. For the other five F-F-A systems, we present the critical scattering length values as a function of mass ratios in Fig.\;\ref{fig2}. As the mass ratio $M/m$ increases, the absolute value of $a_{c}$, when scaled by the van der Waals radius of the F-A system $r_{\text{vdw,FA}}$, tends to become smaller. An interesting observation emerges when comparing the $^{167}$Er,-$^{167}$Er,-$^{6}$Li and $^{171}$Yb,-$^{171}$Yb,-$^{6}$Li systems which have similar mass ratios and $r_{\text{vdw,FA}}$ values. They posses the same critical scattering length. This finding confirms the expected universality of $a_{c},$ hinging on both the mass ratio and $r_{\text{vdw,FA}}$. Figure \ref{fig3} shows the effective potential curves for these systems with $a_{FA}$ fixed at the critical value $a_{c}$. It is shown that the effective potential well tends to become deeper as the mass ratio decreases. This is a manifestation of the fact that a more attractive two-body interaction is needed to form a trimer as the mass ratio approaches the critical mass ratio of $13.6$ at which Efimov effects appear.

\subsection{With Fermion-Fermion interaction}
\begin{figure}[htbp]
\centering
\includegraphics[width=0.5\linewidth]{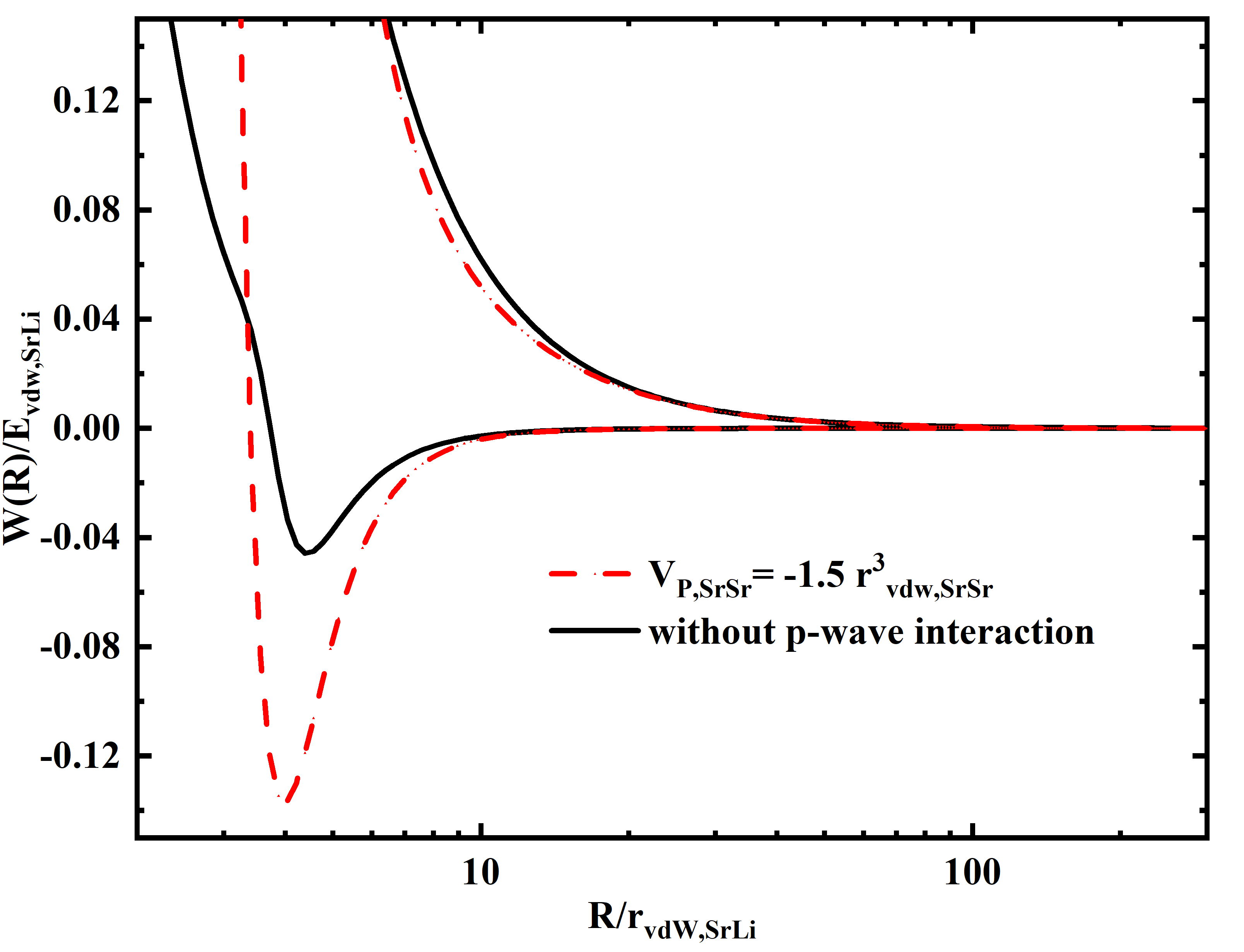}
\caption{Comparison of the $L^{\pi} = 1^{-}$ adiabatic potential curves $W(R)$ for $^{87}$Sr\,-$^{87}$Sr\,-$^{7}$Li with(dash) and without(solid) the $^{87}$Sr-$^{87}$Sr $p$-wave interaction, respectively. The Sr-Li interaction has been fixed at $s$-wave unitary limit($a_{s}\rightarrow\infty$). }
\label{fig4}
\end{figure}

In this section, we consider the model case where the $p$-wave interaction exists between two fermions. We aim to study the influence of $p$-wave fermion-fermion interactions on the critical scattering length value $a_{c}$ where the borromean state appears in the $1^{-}$ symmetry. The $p$-wave attraction between the two fermions is described by the L-J model potential. Similar to the study of F-A interaction, we have adjusted the parameter $\gamma$ to produce the desired scattering volumes $v_p$. For the $^{87}$Sr\,-$^{87}$Sr\,-$^{7}$Li and $^{40}$K\,-$^{40}$K\,-$^{6}$Li systems without fermion-fermion interaction, there isn't any $p$-wave bound state on the negative scattering length side even when the scattering length $a_{\scriptscriptstyle\textsl{FA}}$ is tuned to negative infinite limit. Figure\;\ref{fig4} compares the effective adiabatic potential curves of the $^{87}$Sr\,-$^{87}$Sr\,-$^{7}$Li system for the $L^{\pi} = 1^{-}$ symmetry with and without the fermion-fermion $p$-wave interaction. The $^{87}$Sr and $^{7}$Li atoms' interaction is fixed at $a_{\scriptscriptstyle\textsl{FA}}\rightarrow\infty$. The solid curves represent the potential curves neglecting the $^{87}$Sr\,-$^{87}$Sr $p$-wave interaction, while the dashed lines are the ones with $^{87}$Sr\,-$^{87}$Sr interaction fixed at $v_p =-1.5\, r_{\text{vdw,SrSr}}^{3}$. As can be seen, the effective potential well significantly deepens even with weak fermion-fermion interaction.

\begin{figure}[htbp]
\centering
\includegraphics[width=0.5\linewidth]{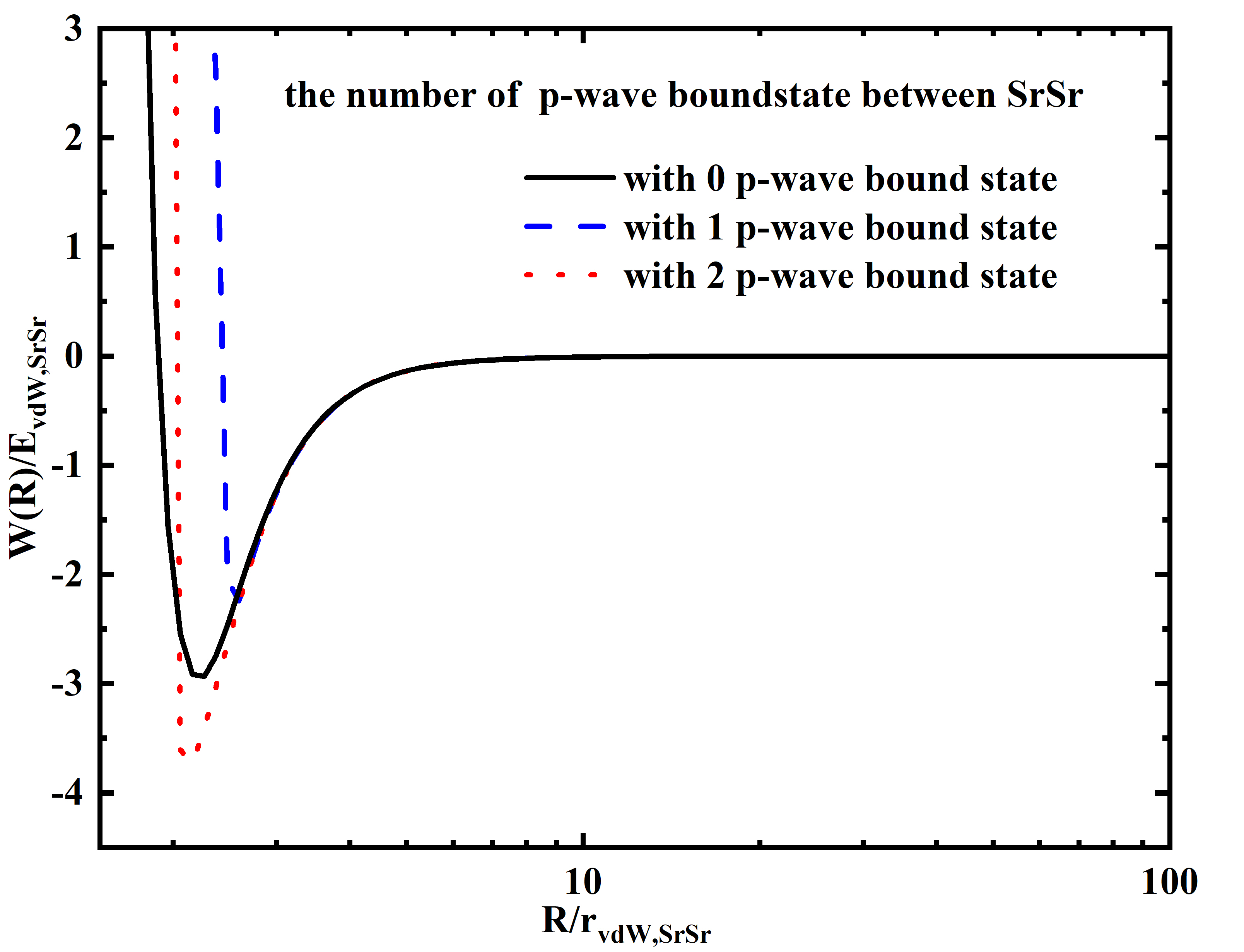}
\caption{The  $L^{\pi} = 1^{-}$ adiabatic potential curves $W(R)$ for $^{87}$Sr\,-$^{87}$Sr\,-$^{7}$Li system with Sr-Li interaction fixed at $s$-wave unitary limit($a_{s}\rightarrow\infty$). The Sr-Sr interaction has been set at $v_{p}=-1.5\, r^3_{\text{vdw,SrSr}}$ for the different $p$-wave pole of the Lennard-Jones potential. }
\label{fig5}
\end{figure}

When the F-A interaction fixed at the $s$-wave unitary, as the fermion-fermion $p$-wave scattering volume $V_{p}$ varied, our calculations show that the $p$-wave borromean state will be created at $v_p =-1.13\, r_{\text{vdw,SrSr}}^{3}$ for $^{87}$Sr\,-$^{87}$Sr\,-$^{7}$Li system. For the $^{40}$K\,-$^{40}$K\,-$^{6}$Li system, the $p$-wave borromean bound state occurs at $v_p =-2.6\, r_{\text{vdw,KK}}^{3}$. Figure \ref{fig4} indicates that this type of borromean state arises due to attractive interactions between two fermions and may be model-dependent for such small scattering volumes. We plot the adiabatic potentials $W(R)$ for $^{87}$Sr\,-$^{87}$Sr\,-$^{7}$Li with various $p$-wave poles of the $^{87}$Sr-$^{87}$Sr interaction to investigate further. Figure \ref{fig5} shows the results, where the $^{87}$Sr\,-$^{7}$Li interaction is set at the $s$-wave unitary limit, and the fermion-fermion interaction is $v_p =-1.5\, r_{\text{vdw,SrSr}}^{3}$ at different $p$-wave poles. The effective adiabatic potentials, derived for various numbers of $^{87}$Sr-$^{87}$Sr two-body $p$-wave bound states, do not converge into a single universal curve. The location of the repulsive barrier shifts with different numbers of $^{87}$Sr-$^{87}$Sr two-body $p$-wave bound states, suggesting that the trimer state relies on the specifics of the $^{87}$Sr-$^{87}$Sr short-range interaction.

\begin{figure}
 \centering
\includegraphics[width=0.5\linewidth]{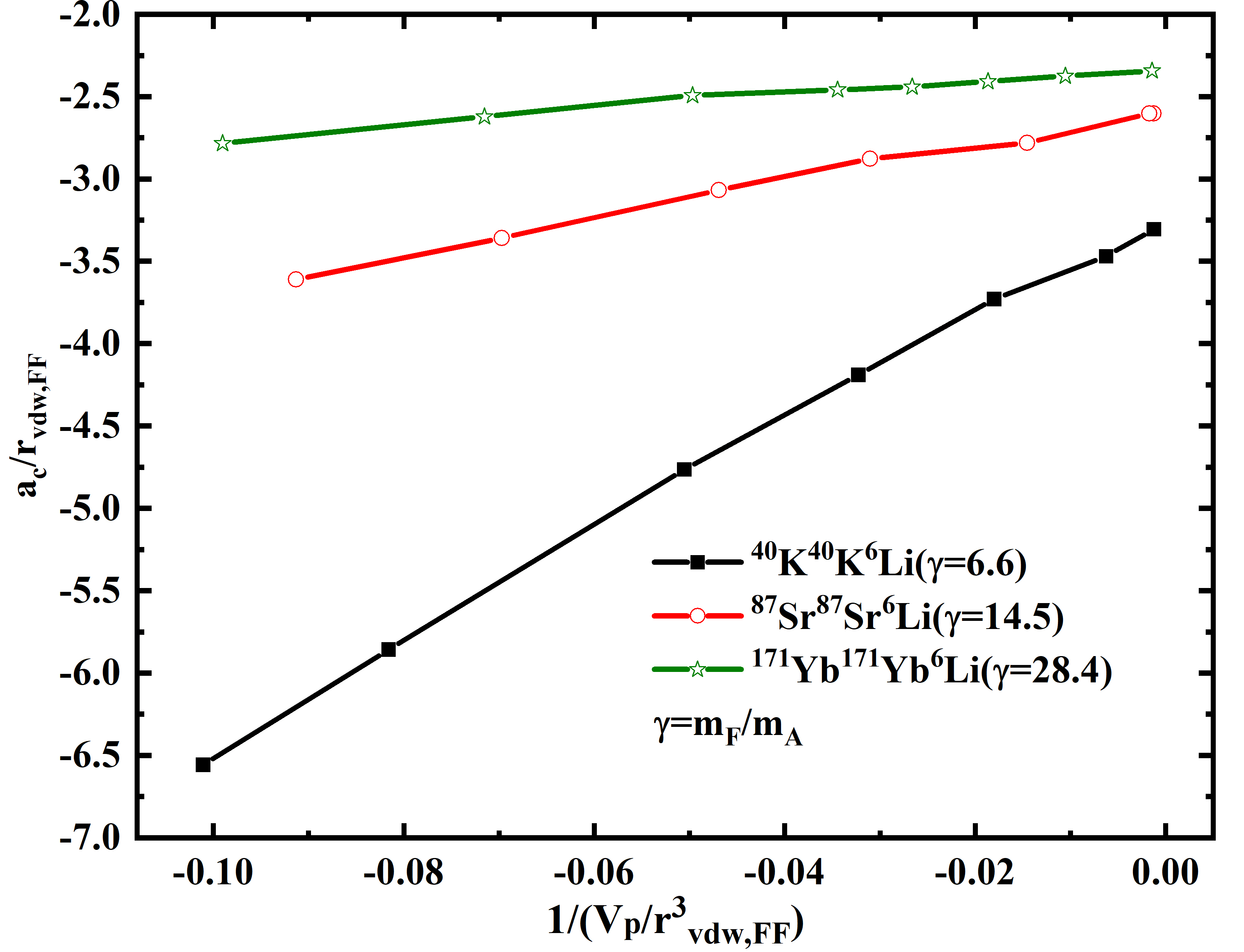}
\caption{(Color online) $a_{ c}$ for $^{40}$K-$^{40}$K-$^{6}$Li, $^{87}$Sr-$^{87}$Sr-$^{7}$Li and $^{171}$Yb-$^{171}$Yb-$^{6}$Li systems, plotted vs the inverse intraspecies $p$-wave scattering volume $1/\left(V_{\scriptscriptstyle{P}}/r_{\text{vdW,FF}}^{\scriptscriptstyle{3}}\right)$. All quantities are scaled by the intraspecies van der Waals length $r_{\text{vdW,FF}}$. }
   \label{fig6}
\end{figure}

Then we explore how the strength of $p$-wave interaction affects the position of the borromean trimer. By fixing the $p$-wave scattering volumes at specific values, we determine the critical scattering length $a_{c}$ of fermion-atom required for the appearance of the borromean state. Figure \;\ref{fig6} displays the results of our calculations. It is evident that stronger fermion-fermion interactions favor the formation of the borromean state, leading to smaller absolute values of $a_{c}$. Notably, this trend is observed in small mass ratio systems such as $^{40}$K-$^{40}$K-$^{6}$Li and $^{87}$Sr-$^{87}$Sr-$^{7}$Li, where $a_{c}$ is strongly dependent on the $p$-wave scattering volume $V_{\scriptscriptstyle{P}}^{\scriptscriptstyle{1/3}}$. In contrast, for large mass ratio systems like $^{171}$Yb-$^{171}$Yb-$^{6}$Li, the $p$-wave interaction plays a less significant role in determining $a_{c}$. This observation aligns with the insensitivity of the Efimov scaling factor to heavy-heavy interactions in systems consisting of two heavy identical bosons and a light atom\;\cite{hwhammer2006}.

\begin{figure}[htbp]
\centering
\includegraphics[width=7cm]{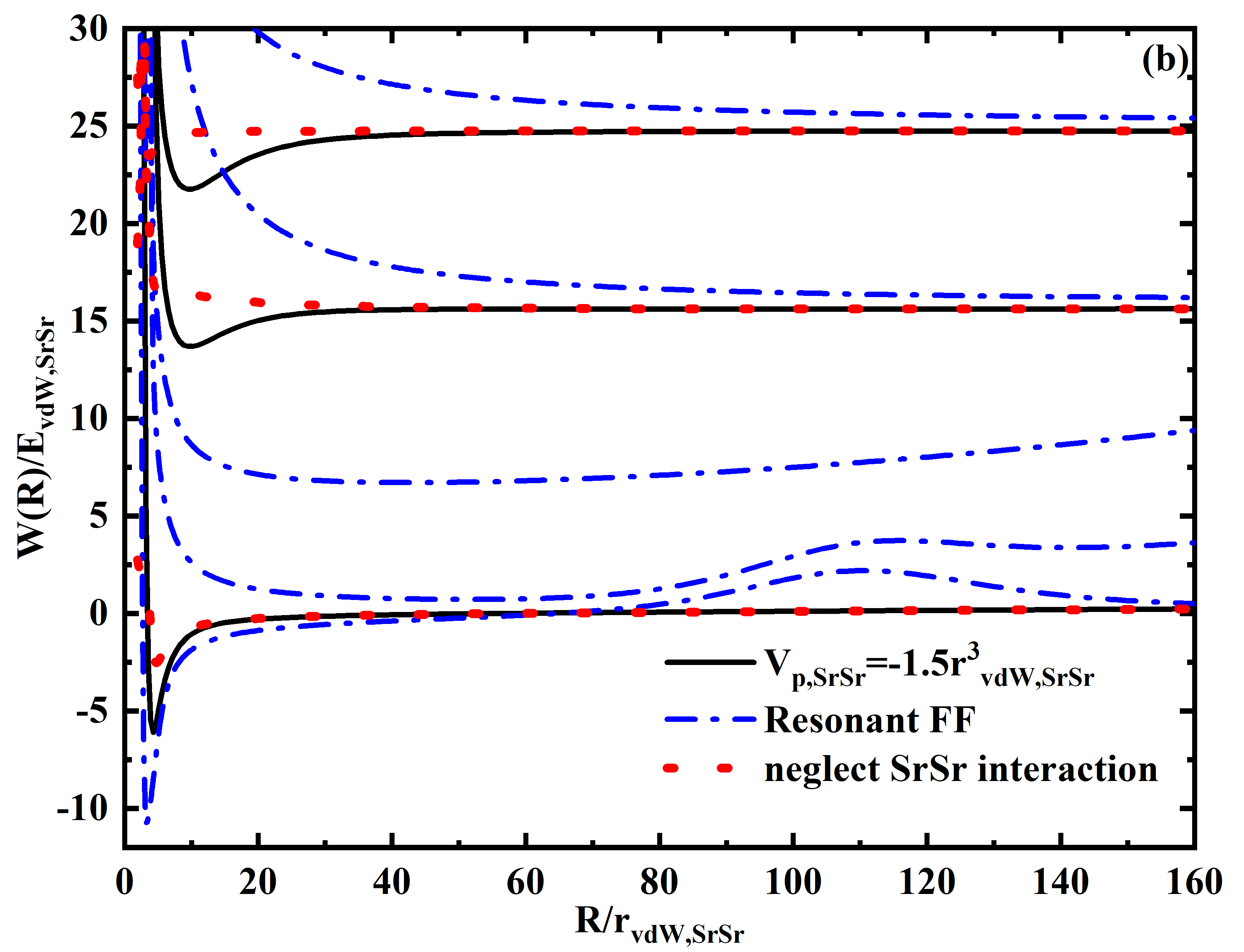}
\caption{(Color online)The adiabatic potential curves $W(R)$ for $^{87}$Sr\,-$^{87}$Sr\,-$^{7}$Li for different strengths of fermion-fermion $p$-wave interaction. The $^{87}$Sr-$^{7}$Li interactions is fixed at the $s$-wave unitary limit($a_{s}\rightarrow\infty$). The dash dot curves are obtained with the $^{87}$Sr-$^{87}$Sr interaction being set at $v_{p}=-\infty$, the solid curves are obtained with the $^{87}$Sr-$^{87}$Sr interaction being set at $v_{p}=- 1.5\, r^{3}_{\text{vdw,SrSr}}$, and the dot lines represent the results neglecting $^{87}$Sr-$^{87}$Sr interaction. These calculations are performed at the first $p$-wave pole of the Lennard-Jones potential. }
\label{fig7}
\end{figure}

Now we focus on the case where the interaction between the two fermions is taken to the $p$-wave unitary limit for the $L^{\pi}=1^{-}$ symmetry. We initially examine the $^{87}$Sr-$^{87}$Sr-$^{7}$Li system, which does not support Efimov states in the absence of $p$-wave interaction between $^{87}$Sr atoms. In Figure\;\ref{fig7}, we present the adiabatic potential curves $2\mu R^2 W(R)$ for this system at varying strengths of $p$-wave interaction. The interaction between $^{87}$Sr and $^{7}$Li is fixed at the $s$-wave unitary limit, and the two $^{87}$Sr atoms interact at the first $p$-wave pole. It is evident that the $p$-wave unitary interaction gives rise to additional potential curves that do not exist in the purely $s$-wave case or in the case with finite $^{87}$Sr-$^{87}$Sr $p$-wave interaction. This intriguing phenomenon aligns with previous observations in two-component fermions of equal mass\;\cite{Chen2022Jan}. Notably, the lowest adiabatic potential curve's asymptotic behavior remains unchanged as the $p$-wave interaction is tuned to infinity, while the potential well becomes significantly stronger.

\begin{figure}[htbp]
\centering
	\subfigure{
	\includegraphics[width=7cm]{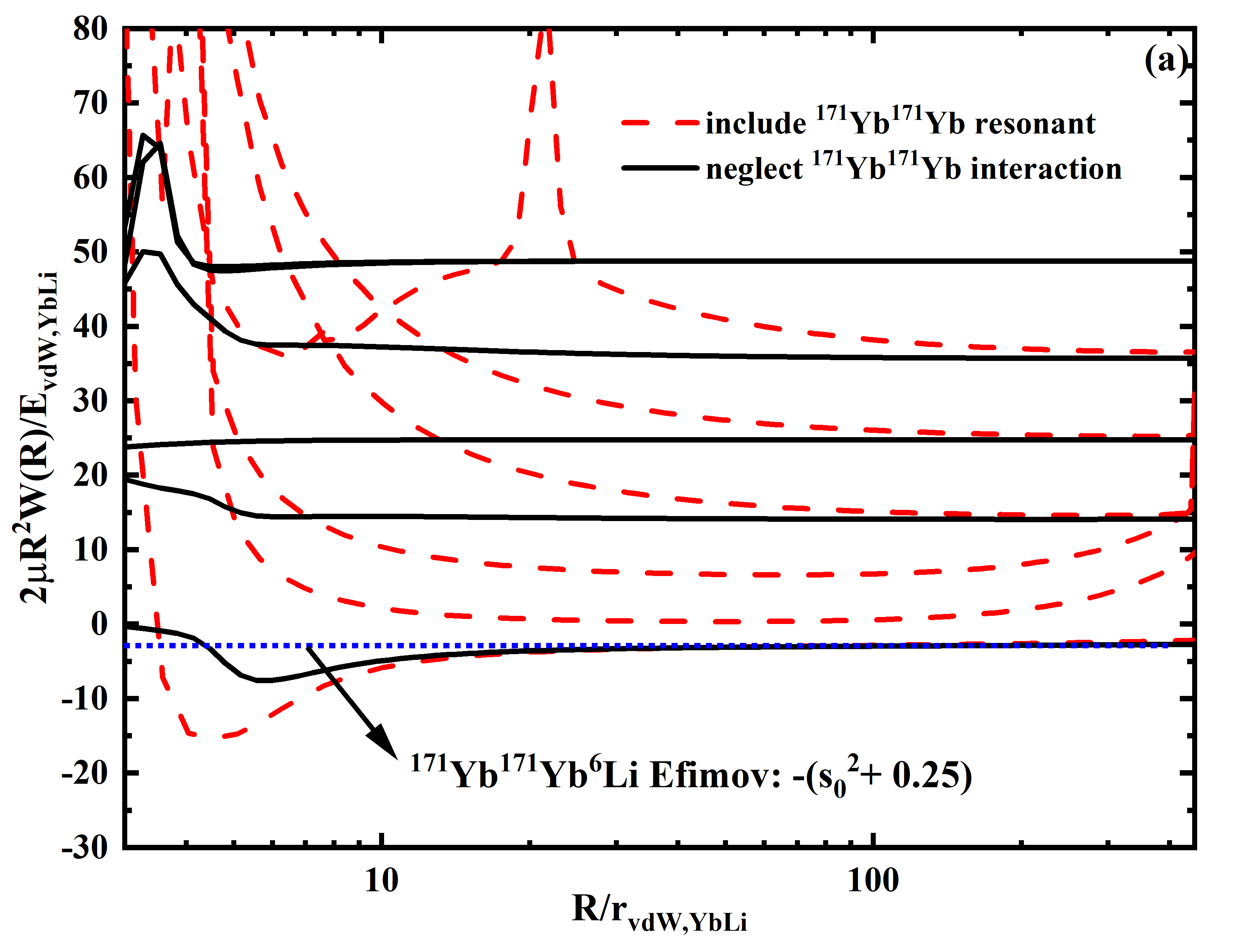}
	\label{fig8a}
}
\subfigure{
	\includegraphics[width=7cm]{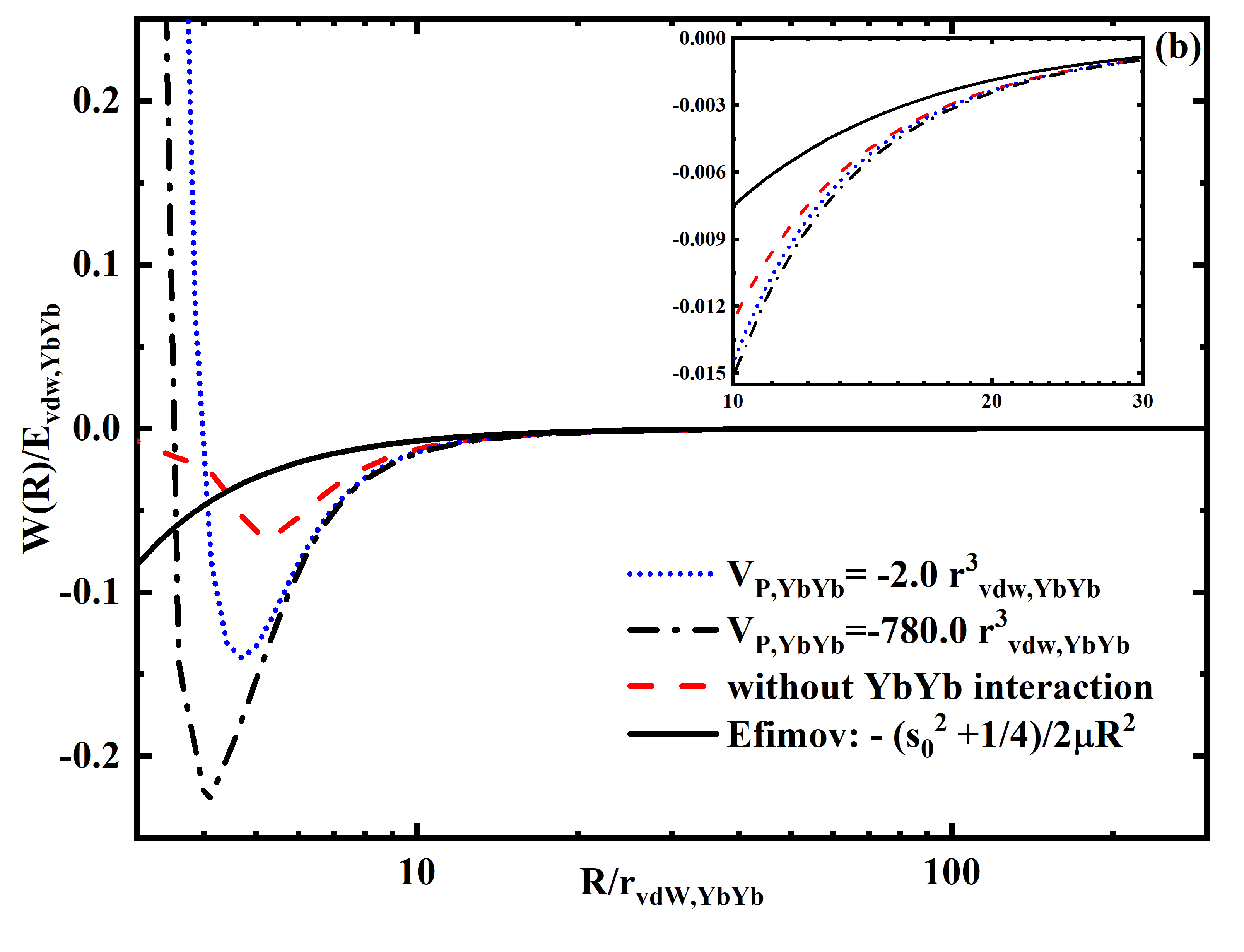}
	\label{fig8b}
}
\caption{The adiabatic potential curves $2\mu R^2 W(R)$ (a) and $W(R)$(b) for $^{171}$Yb\,-$^{171}$Yb\,-$^{6}$Li system with different strengths of fermion-fermion $p$-wave interaction. The $^{171}$Yb\,-$^{6}$Li interactions are fixed at the $s$-wave unitary limit($a_{s}\rightarrow\infty$). (a) Comparison of the adiabatic potential curves $2\mu R^2 W(R)$ for $^{171}$Yb\,-$^{171}$Yb\,-$^{6}$Li system with fermion-fermion $p$-wave resonant interaction and no fermion-fermion interaction,respectively. The blue dot line is the Efimov curve $-( s^2_{0} + 1/4 )/2 \mu R^2$ with $s_{0}=1.628$ (b) Comparison of the adiabatic potential curves $W(R)$ for $^{171}$Yb\,-$^{171}$Yb\,-$^{6}$Li system with different strengths of fermion-fermion $p$-wave interaction.}
\label{fig8}
\end{figure}

\begin{figure}[htbp]
\centering
\includegraphics[width=7cm]{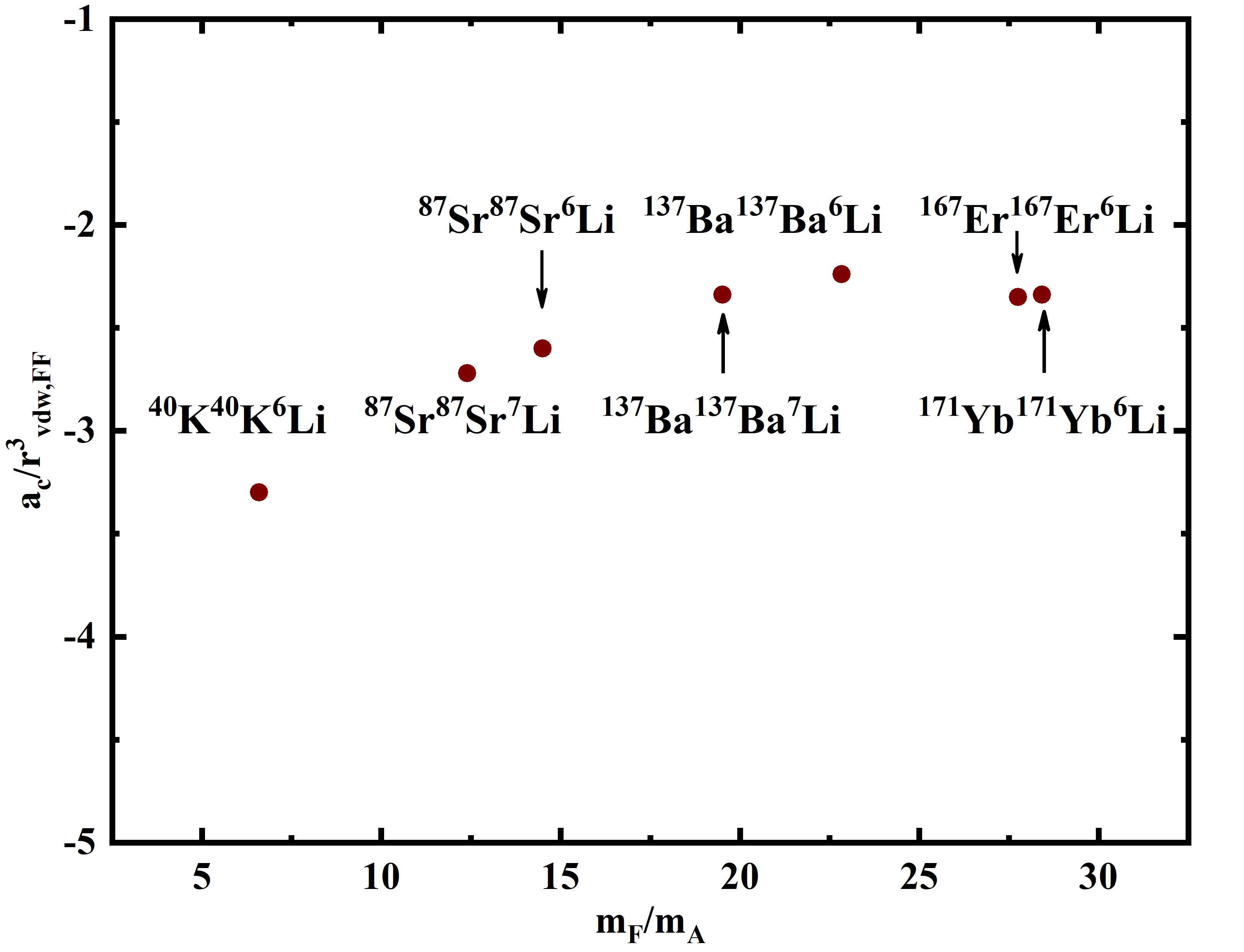}
\caption{(Color online)The values of  $a_{c}$ (scaled by $r_{\text{vdW,FF}}$ ) in FFA systems with resonant fermion-fermion $p$-wave interaction.  }
\label{fig9}
\end{figure}

Figure\;\ref{fig8} displays the $2\mu R^2 W(R)$ and $W(R)$ curves for the $^{171}$Yb-$^{171}$Yb-$^{6}$Li system in $L^{\pi}=1^{-}$ symmetry, including the Efimov curve for comparison. The lowest potential curve (solid line) supports $p$-wave Efimov states when the heavy-light atom interaction approaches the $s$-wave unitary limit in the absence of $p$-wave fermion-fermion interaction. This curve approaches the universal Efimov constant: -($s_{0}^2 + 1/4$) with $s_{0} = 1.628$ throughout the range from $R = 100\, r_{\text{vdw,YbLi}}$ to at least $R=300\, r_{\text{vdw,YbLi}}$ as shown in Fig.\;\ref{fig8a}. The system exhibits additional potential curves arising from the fermion-fermion $p$-wave unitary interaction, similar to the non-Efimovian $^{87}$Sr-$^{87}$Sr-$^{7}$Li system. An interesting finding is that the extent to which the lowest adiabatic potential curve approaches the Efimov curve can be modified by the presence of fermion-fermion $p$-wave interaction, depending on its strength. Figure\;\ref{fig8b} illustrates the differences in the lowest adiabatic potentials for varying strengths of $p$-wave interaction, clearly showing more pronounced deviations in the potential well region with the addition of $p$-wave interaction. This means that the fermion-fermion $p$-wave interaction will affect the critical scattering length at which the $p$-wave borromean state occurs. Figure\;\ref{fig9} shows the critical scattering values $a_{c}$ with resonant fermion-fermion $p$-wave interaction at the first pole. These values fall within the range of $\left[-3.30\, r_{\text{vdW,FF}}, -2.24\, r_{\text{vdW,FF}}\right]$ after being scaled by the characteristic length scale $r_{\text{vdw,FF}}$.

\section{Conclusions}
To summarize, we examined the impact of $p$-wave fermion-fermion interactions on Efimov physics in three-particle systems comprised of two heavy fermions with identical masses $M$ and a lighter particle with mass $m$, where the mass ratios were greater than 13.6.
Our analysis employed a single-channel Lennard-Jones interaction featuring long-range two-body van der Waals potentials. We focused primarily on the borromean regime, where the ground-state trimer with $L^{\pi}=1^{-}$ symmetry exists in the absence of dimers. The critical value of the interspecies $s$-wave scattering length $a_{c}$ at which the borromean state appears for several two-component particle systems is studied. We find that with negligible $p$-wave fermion-fermion interaction, $a_{c}$ is universally determined by the van der Waals radius and mass ratio $M/m$. Evidence for the universality of the critical value has emerged from our tests of different $s$-wave poles of two-body Lennard-Jones potentials. However, van der Waals universality no longer holds in the presence of p-wave fermion-fermion interactions. Our calculation shows that the critical interspecies scattering length $a_{c}$ depend on the short details of fermion-fermion p-wave interaction. Notably, results obtained at the first $p$-wave pole of fermion-fermion interaction may vary in real systems featuring multiple two-body bound states. The formation of the borromean state is promoted in the presence of attractive interactions between the two fermions. We also demonstrated that $p$-wave Efimov effects persist even at the $p$-wave unitary limit of the fermion-fermion interaction.

\section{Acknowledgments}
We thank C. H. Greene for helpful discussions. Hui-Li Han was
supported by the National Natural Science Foundation of China under Grants Nos. 12374235 and 12374235. Ting-Yun Shi was supported by National Natural Science Foundation of China under Grants No12274423.
All the calculations are done on the APM-Theoretical Computing Cluster(APM-TCC). Any data that support the findings of this study are included within the article.


%

\end{document}